# Sequential phase locking scheme for filled aperture intensity coherent combination of beam arrays

**Arno Klenke,**[1,2,*] **Michael Müller,**[1] **Henning Stark,**[1] **Andreas Tünnermann**[1,2,3] **and Jens Limpert**[1,2,3]

[1] *Institute of Applied Physics, Abbe Center of Photonics, Friedrich-Schiller-Universität Jena, Albert-Einstein-Straße 15, 07745 Jena, Germany*
[2] *Helmholtz-Institute Jena, Fröbelstieg 3, 07743 Jena, Germany*
[3] *Fraunhofer Institute for Applied Optics and Precision Engineering, Albert-Einstein-Str. 7, 07745 Jena, Germany*
*\* a.klenke@gsi.de*

**Abstract:** We present a novel phase locking scheme for the coherent combination of beam arrays in the filled aperture configuration. Employing a phase dithering mechanism for the different beams similar to LOCSET, dithering frequencies for sequential combination steps are reused. By applying an additional phase alternating scheme, this allows to use standard synchronized multichannel lock-in electronics for phase locking a large number of channels even when the frequency bandwidth of the employed phase actuators is limited.

## 1. Introduction

The demand for laser systems with ever increasing performance figures has led to the development of advanced laser concepts. One approach that has been applied especially to fiber lasers over the recent years, is the parallelization of the amplification process by using many laser amplifiers in conjunction with subsequent coherent beam combination [1]. This concept is applied successfully to sources reaching from continuous-wave to femtosecond regimes. One of the most important building blocks of such systems is the phase locking of the beams to maintain constructive interference. Different techniques can be employed, depending on the combination approach used. The Hänsch-Couillaud technique [2] can be applied to the polarization combination approach, since it can measure the polarization state of a combined beam created from two orthogonal linearly polarized sub-beams. This measurement allows to derive the phase difference between two beams, including the sign. The technique can also be employed in a cascaded setup to combine more than two beams [3]. However, one polarization detector is needed for every combination setup ($N_{total}$-1 for a complete setup, with $N_{total}$ being the total number of channels). Another combination approach is to use partially reflective beam-splitters to combine beams in the same polarization state. In this case, the combined beam does not contain any directly recoverable information about the sign of the phase difference. Therefore, small phase modulations have to be applied to the different phase actuators that translate to small intensity fluctuations of the combined beams. An error signal with the proper sign can then be deduced from the combined signal in conjunction with the known applied phase modulations. Two representatives of this approach are LOCSET [4] and SPGD [5]. In most cases, both techniques are implemented by just using a single intensity detector at the output, even for a large number of channels $N_{total}$.

In this paper, we present a novel sequential phase locking scheme that depends on a multi-detector configuration for intensity beam combination. It is especially useful for systems where the actuator bandwidth limits the number of channels $N_{total}$, for example when using piezo based actuators.

## 2. Beam combination setup

The presented scheme requires coherent combination setups where the beams are coherently added one-by-one into a combined beam, which excludes alternative combination schemes such as tiled-aperture coherent combination. It especially works well for the combination of arrays comprising equidistant and parallel beams into a single output beam.

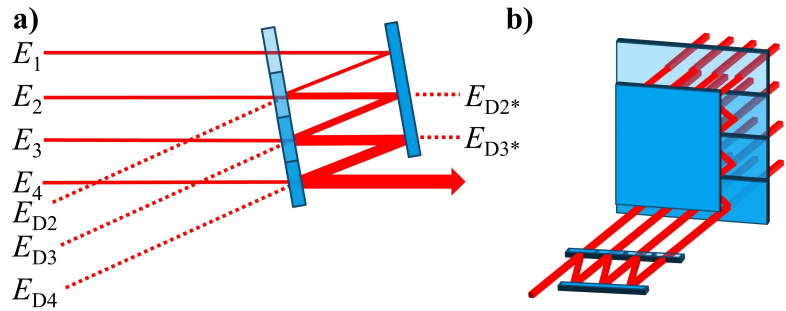

Fig. 1. a) Beam combination scheme for a one-dimensional beam array for four beams ($E_1$-$E_4$). The dashed lines are the non-combining parts of the beams that form the rejection port. Possible photodiode detector positions for the combined powers are in this rejection port ($E_{D2}$-$E_{D4}$) and behind the HR-mirror ($E_{D2*}$-$E_{D3*}$, $E_{D4*}$ could be retrieved by using a fraction of the combined beam), b) Combination scheme for a two-dimensional beam array

This can be realized for a full-aperture configuration (all beams overlap in the near-field and the far-field) with a segmented mirror splitter/combiner [6] consisting of an optical element with zones of different reflectivity parallel to a highly reflective mirror (see Fig. 1(a)). The component can act as a beam splitter as well as a combiner. For splitting an incident beam into beams with equal power or for combining such beams, the reflectivity for the $N$-th zone is chosen as $(N-1)/N_{total}$. The first beam, the top one in Fig. 1(a), propagates in a zig-zag path and interferes at every segment of the element with the next beam, therefore adding power if constructive interference occurs. This splitter/combiner can also be used twice for a two-dimensional configuration (see Fig. 1(b)). The setup offers access to the intensity of the combined beam and the non-combining parts at every combination step in the leakage of the HR mirror and at the back reflections of the partially reflective element, respectively.

## 3. Phase locking

The major phase locking concept when employing intensity combination is LOCSET, which works by adding small phase modulations with unique frequencies to each channel. These frequencies need to be separated by the locking bandwidth in order to prevent crosstalk and provide sufficient bandwidth for each channel. However, phase actuators have a limited bandwidth themselves, thus limiting the usable number of channels. This especially holds true for piezo-based actuators that can provide larger piston changes. These kind of actuators are not required for the phase locking of CW lasers or lasers emitting pulses with nanosecond duration, where an actuator providing a phase change between 0 and 2π is sufficient. Actuators like electro-optical modulators with GHz bandwidths are available for such setups. However, for the combination of ultrashort pulses the path lengths need to be matched precisely to a few wavelengths in order to provide a high combination efficiency [7]. The actuator should be able to compensate for dynamic path length fluctuations during operation, which typically requires a piston change of multiple µm. Alternatively to LOCSET, time dependent single frequency modulation schemes [8] or combined schemes like CDMA [9] can be employed. Recently, a modified LOCSET scheme based on multiple detectors was demonstrated that separates the channels into multiple groups, first phase locking all the beams in each group and then the resulting combined beams themselves [10]. Therefore, the same modulation frequencies can be reused in each group. Still, in a two-dimensional setup with $N_{total} \times M_{total}$ beams, $N_{total}+M_{total}$ different frequency bands are required.

A straightforward idea is to reuse the same modulation frequency band in sequential combination steps and to employ a photodiode after every combination step (e.g. by using the residual light of the beams behind the HR mirror (transmission $T_{HR}$) in Fig. 1(a)) to measure the respective combined signal. However, assuming no phase modulation for the first channel and equal ones for the other channels, the following calculation shows that the measured intensity signal of the electric field at the detector for the $N$-th channel $E_{D\,N^*}$ (and the corresponding intensity $I_{D\,N^*}$) contains the same contribution for the phase of channel $N$ as for the previous channels. In this calculation, equal electric field amplitudes $E_0$ are assumed for all beams and a phase modulation function $\phi_{mod}(t)$ is applied to each beam except the first one. Furthermore, a phase error $\phi_i$ relative to the first beam is added. $n = N-1$ is the number of previous channels, $E_{comb\,N}$ the combined electric field of the $N$ channels and all terms not depending on the modulation function were discarded:

$$E_{comb\,N} = \sum_{r=1}^{N} \sqrt{\frac{1}{N}} E_r, \quad E_1 = E_0, \quad E_r = E_0 \exp\left(i\left(\phi_{mod}(t) + \phi_r\right)\right)$$

$$E_{D\,N^*} = \sqrt{T_{HR}} E_{comb\,N} = \sqrt{T_{HR}} \left(\sqrt{\frac{1}{N}} E_N + \sqrt{\frac{1}{N}} \sum_{r=1}^{n} E_r\right)$$

$$I_{DN^*} \sim |E_{DN^*}|^2$$

$$= T_{HR} \begin{pmatrix} \dfrac{1}{N}|E_N|^2 + \dfrac{1}{N}\sum_{r=1,s=1}^{n}\left(E_r^* E_s + cc\right) + \\ \dfrac{1}{N}\sum_{r=1}^{n}\left(E_N^* E_r + cc\right) \end{pmatrix}$$

$$= T_{HR} \begin{pmatrix} const + \dfrac{1}{N}\sum_{r=2}^{n} 2|E_0|^2 \cos\left(\phi_{\mathrm{mod}}(t)+\phi_r\right) + \\ \dfrac{1}{N} 2|E_0|^2 \cos\left(\phi_{\mathrm{mod}}(t)+\phi_N\right) \end{pmatrix}$$

(1)

By multiplying this detector signal with the modulation function and integrating over time (realized with a low pass circuit), an error signal $S_N$ can be generated. If a phase dither function $\phi_{\mathrm{mod}}(t)=\beta\sin(\omega_{\mathrm{mod}}t)$ is assumed, this error signal can be calculated by using the Fourier series expansion of the cosine function [11]:

$$S_N \underset{\tau \gg \frac{1}{\omega_{\mathrm{mod}}}}{\sim} \frac{1}{\tau}\int_0^\tau I_D \sin(\omega_{\mathrm{mod}} t)\,dt = -\frac{2}{N}|E_0|^2 J_1(\beta)\left(\sum_{r=2}^{n}\sin(\phi_r)+\sin(\phi_N)\right)$$

(2)

This result is not viable for phase locking which can be demonstrated with a simple example for a four channel system.

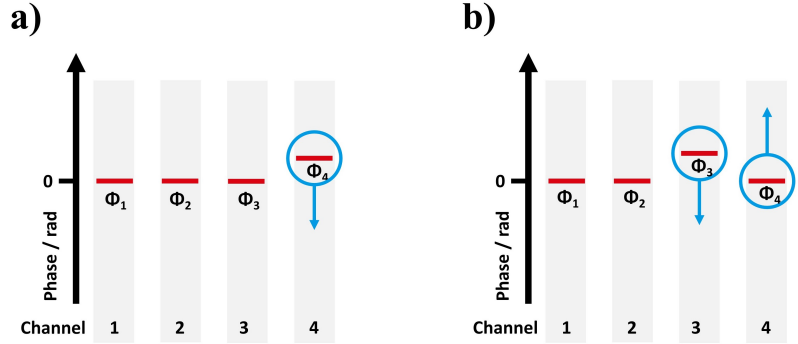

Fig. 2. Example phase offsets (red) for a four channel system and the required movement of the phase actuators (blue) for sequential phase locking: (a) the fourth channel has a positive phase offset and the actuator is supposed to decrease it to optimize for best combination efficiency and (b) the third channel has a positive phase offset. The actuators are supposed to decrease the phase offset of the third channel, but increase it for the fourth channel to adapt to the combined beam of the first three channels.

In Fig 2(a), the situation is shown with the first three beams being in phase, while the fourth and last one has a phase offset of $\phi_4$. According to the error signal defined above, this results in a negative error signal for channel 4, meaning the corresponding phase actuator decreases $\phi_4$.

Equivalently, if a phase offset of $\phi_3$ is applied to channel 3 (as depicted in Fig 2(b)), this introduces a negative error signal and an according phase actuator reaction in channel 3. As can be deduced from the previous equation, this also induces a negative error signal in channel 4. However, this error signal is required to be positive so that the phase of this channel is increased by the phase actuator and adjusted to the, on average, positive phase offset of the combined beam of the preceding three channels. Consequently, due to this error signal having the wrong sign and, therefore, the phase actuator moving in the wrong direction, a phase correction using this approach is impossible. It should be noted, however, that this detector configuration would work in the LOCSET scheme, because there the error signal only includes the phase difference to the unmodulated channel and no contribution of the other phase offsets.

This result can be changed by placing the signal detectors in the respective rejection ports (for the $N$-th channel electric field $E_{D\,N}$ and intensity $I_{D\,N}$). The two beams (the combined beam of the $n = N-1$ previous channels and the $N$-th beam) that interfere in the rejection port see an additional relative $\pi$ phase difference in contrast to the beams combined on the other side of the optical element. Calculating the detector signal again results in:

$$E_{\text{comb } N} = \sum_{r=1}^{N} \sqrt{\frac{1}{N}} E_r, \quad E_1 = E_0, \quad E_r = E_0 \exp\left(i\left(\phi_{\text{mod}}(t) + \phi_r\right)\right)$$

$$E_{DN} = \sqrt{\frac{N-1}{N}} E_N - \sqrt{\frac{1}{N}} \sum_{r=1}^{n} \sqrt{\frac{1}{N-1}} E_r$$

$$I_{DN} \sim |E_{DN}|^2 = \left( \begin{array}{l} \frac{N-1}{N} |E_N|^2 + \frac{1}{N(N-1)} \sum_{r=1,s=1}^{n} \left(E_r^* E_s + cc\right) \\ -\frac{1}{N} \sum_{r=1}^{n} \left(E_N^* E_r + cc\right) \end{array} \right)$$

$$= \text{const} + \left( \begin{array}{l} \frac{1}{N(N-1)} \sum_{r=2}^{n} 2|E_0|^2 \cos\left(\phi_{\text{mod}}(t) + \phi_r\right) \\ -\frac{1}{N} 2|E_0|^2 \cos\left(\phi_{\text{mod}}(t) + \phi_N\right) \end{array} \right)$$

(3)

And the error signal can be calculated accordingly (the error signal was additionally negated to make the comparison with the previous case easier):

$$S_N \underset{\tau \gg \frac{1}{\omega_{\text{mod}}}}{\sim} -\frac{1}{\tau} \int_0^\tau I_{DN} \sin(\omega_{\text{mod}} t) dt = \frac{2}{N} |E_0|^2 J_1(\beta) \left( \begin{array}{l} \frac{1}{N-1} \sum_{r=2}^{n} \sin(\phi_r) \\ -\sin(\phi_N) \end{array} \right)$$

(4)

The required sign difference between the signal term created by the $N$-th channel and by the $n$ previous channels is present in this equation and transfers to the error signal accordingly. In the example mentioned previously in Fig. 2, the phase actuator for the $N$-th channel would now move into the right direction. However, the modulation contrast for the $N$-th channel drops with an increasing number of channels. Therefore, an increase of the dither amplitude might be necessary.

In order to counter this issue, the absolute phases of the applied phase dither functions can be changed. In an experimental setup, this can be achieved by using synchronized multi-channel lock-in electronics with adjustable offsets for the phase modulation and demodulation stages. In the following, this approach is explored in a configuration where the first beam $E_1$ is not phase modulated, followed by subsequent alternating phase modulations with 0 ($E_{+r}$) and $\pi$ phase modulation offset ($E_{-r}$):

**Table 1. Sample channel configuration for the phase alternating scheme**

| Channel | 1 | 2 | 3 | 4 | 5 | … |
|---|---|---|---|---|---|---|
| Phase modulation offset | not modulated | 0 | $\pi$ | 0 | $\pi$ | … |
| Electric field | $E_1$ | $E_{+2}$ | $E_{-3}$ | $E_{+4}$ | $E_{-5}$ | … |

In order to analyze the resulting detector and error signal for channel $N$, the cases where this channel has a 0 phase offset and the case where it has a $\pi$ phase offset have to be considered separately. Those will be referred to as '+' channels and '−' channels in the following. In the first case, for channel $N$ there are $N/2-1$ previous '+' channels and the same number of '−' channels. The detector signal is then:

$$E_1 = E_0, E_{+r} = E_0 \exp\left(i\left(\phi_{\mathrm{mod}}(t) + \phi_{+r}\right)\right), E_{-r} = E_0 \exp\left(i\left(-\phi_{\mathrm{mod}}(t) + \phi_{-r}\right)\right)$$

$$E_{+N} = E_0 \exp\left(i\left(\phi_{\mathrm{mod}}(t) + \phi_{+N}\right)\right)$$

$$E_n = \sqrt{\frac{1}{n}}\left(E_1 + \sum_r E_{+r} + \sum_r E_{-r}\right)$$

$$|E_n|^2 = \frac{1}{n}\left(\begin{array}{l}|E_0|^2 + \left|\sum_r E_{+r}\right|^2 + \left|\sum_i E_{-r}\right|^2 + \left(\left(\sum_r E^*_{+r}\right)\left(\sum_r E_{-r}\right)\right) \\ + \left(E_0^* \sum_r E_{+r} + cc\right) + \left(E_0^* \sum_r E_{-r} + cc\right)\end{array}\right)$$

$$= const + \frac{1}{n}\left(\begin{array}{l}\left(E_0^* \sum_r E_{+r} + cc\right) + \left(E_0^* \sum_r E_{-r} + cc\right) \\ + 2\sum_r \sum_s |E_{+r}||E_{-s}|\cos\left(2\phi_{\mathrm{mod}}(t) + \phi_{+r} - \phi_{-s}\right)\end{array}\right)$$

$$= const + \frac{1}{n}|E_0|^2 \left(\begin{array}{l} 2\left(\sum_r \cos\left(\phi_{\mathrm{mod}}(t) + \phi_{+r}\right)\right) + 2\left(\sum_r \cos\left(-\phi_{\mathrm{mod}}(t) + \phi_{-r}\right)\right) \\ + 2\sum_r \sum_s \cos\left(2\phi_{\mathrm{mod}}(t) + \phi_{+r} - \phi_{-s}\right)\end{array}\right)$$

$$I_{DN} \sim |E_{DN}|^2 = \left|\sqrt{\frac{N-1}{N}}E_{+N} - \sqrt{\frac{1}{N}}E_n\right|^2$$

$$= const + \frac{1}{N}|E_n|^2 - \frac{\sqrt{N-1}}{N}\sqrt{\frac{1}{n}}|E_0|^2 \left(\begin{array}{l} 2\cos\left(\phi_{\mathrm{mod}}(t) + \phi_{+N}\right) \\ + 2\left(\sum_r \cos\left(2\phi_{\mathrm{mod}}(t) + \phi_{+N} - \phi_{-r}\right)\right)\end{array}\right) \quad (5)$$

To simplify this equation, a control loop analysis can be done for small phase errors and small phase modulations as in [4]. To do this, the cosine function is replaced by its Taylor series expansion and only terms with a linear dependency on the modulation function are considered:

$$I_{DN} \sim const + \frac{1}{N}\frac{1}{N-1}|E_0|^2 \begin{pmatrix} -2\left(\sum_r \phi_{mod}(t)\phi_{+r}\right) + 2\left(\sum_r \phi_{mod}(t)\phi_{-r}\right) \\ -4\sum_r\sum_s (\phi_{+r} - \phi_{-s})\phi_{mod}(t) \end{pmatrix}$$

$$-\frac{1}{N}|E_0|^2 \left(-2\phi_{mod}(t)\phi_{+N} - 4\sum_r (\phi_{+N} - \phi_{-r})\phi_{mod}(t)\right)$$

$$\sim const - \frac{1}{N}\frac{1}{N-1}|E_0|^2 \left((2(N-1))\left(\sum_r \phi_{mod}(t)\phi_{+r}\right) - (2(N-1))\left(\sum_r \phi_{mod}(t)\phi_{-r}\right)\right)$$

$$+\frac{1}{N}|E_0|^2 \left(2\phi_{mod}(t)\phi_{+N} + 4\left(\frac{N}{2}-1\right)\phi_{+N}\phi_{mod}(t) - 4\sum_r \phi_{mod}(t)\phi_{-r}\right)$$

$$\sim const - \frac{1}{N}|E_0|^2 \left(2\left(\sum_r \phi_{+r}\right) + 2\left(\sum_r \phi_{-r}\right) - 2(N-1)\phi_{+N}\right)\phi_{mod}(t)$$

$$S_{+N} \underset{\tau \gg \frac{1}{\omega_{mod}}}{\sim} \frac{1}{N}|E_0|^2 \beta^2 \left(\left(\sum_r \phi_{+r}\right) + \left(\sum_r \phi_{-r}\right) - (N-1)\phi_{+N}\right)$$

(6)

As can be seen, the sign of the error signal shows the correct values for the previous '+' and '–' channels, as well as for the considered channel N. Additionally, the signal amplitude stays basically constant with an increasing channel number N. The same calculation can also be done for N being a channel with π phase offset. In this case, there are *(N-1)/2* previous '+' channels and *(N-3)/2* '–' channels and the phase modulation offset needs to be considered for the demodulation step. The result for the detector signal and error signal is very similar to the case above, as expected:

$$|E_D|^2 = \left|\sqrt{\frac{N-1}{N}}E_{-N} - \sqrt{\frac{1}{N}}E_n\right|^2$$

$$= const + \frac{1}{N}|E_n|^2 - \frac{\sqrt{N-1}}{N}\sqrt{\frac{1}{n}}|E_0|^2 \begin{pmatrix} 2\cos(-\phi_{mod}(t) + \phi_{-N}) \\ +2\left(\sum_r \cos(-2\phi_{mod}(t) + \phi_{-N} - \phi_{+r})\right) \end{pmatrix}$$

$$|E_D|^2 \sim const + \frac{1}{N}\frac{1}{N-1}|E_0|^2 \begin{pmatrix} -2\left(\sum_r \phi_{mod}(t)\phi_{+r}\right) + 2\left(\sum_r \phi_{mod}(t)\phi_{-r}\right) \\ -4\sum_r\sum_s (\phi_{+r} - \phi_{-s})\phi_{mod}(t) \end{pmatrix}$$

$$-\frac{1}{N}|E_0|^2 \left(2\phi_{mod}(t)\phi_{-N} + 4\sum_r (\phi_{-N} - \phi_{+r})\phi_{mod}(t)\right)$$

$$\sim const - \frac{1}{N}|E_0|^2 \left(-\frac{2N}{N-1}\left(\sum_r \phi_{+r}\right) - \frac{2N}{N-1}\left(\sum_r \phi_{-r}\right) + 2N\phi_{-N}\right)\phi_{mod}(t)$$

$$S_{-N} \underset{\tau \gg \frac{1}{\omega_{mod}}}{\sim} \frac{1}{N}|E_0|^2 \beta^2 \left(\frac{N}{N-1}\left(\sum_r \phi_{+r}\right) + \frac{N}{N-1}\left(\sum_r \phi_{-r}\right) - N\phi_{-N}\right)$$

(7)

Therefore, the presented configuration of N channels in one dimension is viable. Of course, the sequence of '+' and '–' channels can be swapped and the error signals can be calculated accordingly by adapting the equations above.

If we assume that this dimension is one column of a two-dimensional beam array (see Fig. 1(b)), a second modulation frequency has to be introduced that modulates the first actuators of each additional column, which will be explained below

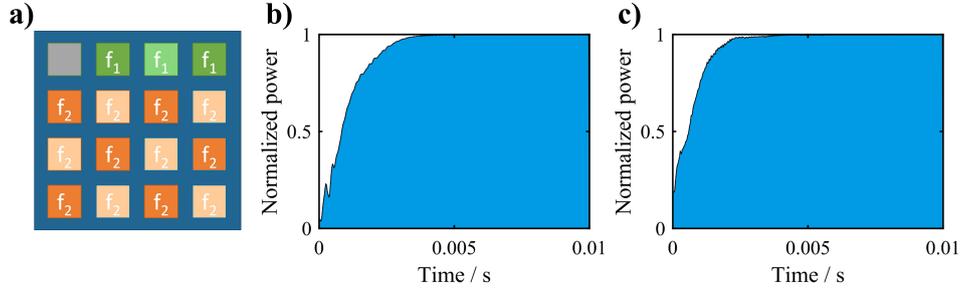

Fig. 3. a) Example configuration of applied phase dither for a two-dimensional 4x4 actuator array with two different frequencies $f_1$ and $f_2$ and a phase offset alternating scheme indicated by the darker and lighter colors. b) Time dependence of the normalized combined power for the run with the highest convergence time of 1000 numerical simulations starting with random initial phases for each beam). c) Simulation of the LOCSET scheme for the same beam array combination with different phase dither frequencies for each actuator.

An example configuration is shown in Fig. 3(a). If the first actuator of the first column is not modulated, a frequency $f_1$ with alternating phase offsets can be applied to the first actuators of each following column and a frequency $f_2$ to the other actuators. The reason for this additional frequency $f_1$ is that in the second dimension combination steps, we do not coherently add beams with well-defined phase modulations one-by-one, but already combined beams from the first dimension. Therefore, the assumptions in the presented calculations are not valid in this case. However, by adding this second frequency the equations above can then be applied for the first combination step that combines the beams from each column (using modulation frequency $f_2$), followed by a second combination step that combines this row of combined beams into the single output beam (using modulation frequency $f_1$). Such a setup has been simulated for a 4x4 beam array using phase dithers of $f_1$=6 kHz and $f_2$=4 kHz with dither amplitudes of $\pi/70$. The numerical model implements the required demodulation of the combined signals to calculate an error signal for each beam and modifies the beam phases accordingly through a simulated regulator. Figure 3(b) shows the time dependent normalized combined power for the run with the highest convergence time out of 1000 runs with random initial phases for each beam. For comparison, in Fig. 3(c), the LOCSET scheme is simulated for phase dither frequencies starting from 2 kHz, with subsequent actuators spaced out by 1 kHz for a maximum frequency of 17 kHz. As can be seen from the two plots, both schemes show a comparable time dependent convergence to the maximum power, even though the LOCSET scheme requires a much higher maximum actuator frequency.

### 4. Conclusion

In conclusion, we have presented a concept for sequential phase locking for coherent beam combining. It works with filled-aperture intensity beam combination setups where the beams are coherently added one-by-one and uses one detector for every combination step. With a segmented-mirror-splitter/combiner and detector arrays, this allows compact beam array combination setups that can be scaled up to a large number of parallel channels even for phase actuators with limited bandwidth. By placing the detector arrays in the rejection port of the combination, an error signal can be generated that shows the right behavior to optimize the system to minimum phase differences and, thus, to maximum combination efficiency. The scheme can be employed by using all actuators in-phase or by using phase offset alternations of 0 and $\pi$. The in-phase version might require an increase of the phase dither amplitude to keep the error signal strength constant for subsequent channels while the phase offset alternating version only requires that the phase dither amplitude is maintained.

**Funding**

European Research Council (ERC) (670557), "MIMAS"; Fraunhofer research cluster "Advanced Photon Sources".

**Acknowledgments**

M.M. acknowledges financial support by the Carl-Zeiss-Stiftung.